\documentstyle[epsfig,epsf]{article}
\begin{document}

\begin{center}
{\Large {\bf Searches for Magnetic Monopoles, Nuclearites and Q-balls}}
\end{center}

\vskip .7 cm

\begin{center}
G. GIACOMELLI, S. MANZOOR, E. MEDINACELI and L. PATRIZII\par~\par
{\it Dept of Physics, Univ. of Bologna and INFN, \\
v.le C. Berti Pichat 6/2, Bologna, I-40127, Italy\\} 

E-mail: giacomelli@bo.infn.it , patrizii@bo.infn.it,
medinaceli@bo.infn.it, manzoor@bo.infn.it

\par~\par

Lectures at the 2$^{nd}$ School on Cosmic Rays and Astrophysics,\\
Puebla, Mexico, August - September 2006.

\end{center}

{\bf Abstract.} \normalsize The searches  for classical Dirac Magnetic
Monopoles (MMs) at accelerators,  
for GUT superheavy MMs in the penetrating cosmic radiation 
and  for Intermediate Mass MMs at high altitudes are discussed. Also
the searches for 
nuclearites and Q-balls are considered. 
\vspace{2mm}

\section{Introduction}\label{sec:intro}
Though the concept of magnetic monopoles (MMs) goes back even to the origin of
magnetism, it is only since 1931 that systematic searches have been performed.
In 1931 Dirac introduced the MM in order to explain the 
quantization of the electric charge~\cite{dirac}. He established the
relation between the  elementary electric charge $e$ and a basic
magnetic charge $g$:  $~~eg=n\hbar c/2= ng_{D}$, \hspace{3mm}
where $n$ is an integer, $n=1,2,..$; $g_D=\hbar c/2e = 68.5 e$ is the
unit Dirac charge. The existence of magnetic charges and of magnetic
currents would symmetrize in form  Maxwell's equations, but there
would be a numerical asymmetry since $e \neq g$ (but the couplings 
could be energy dependent and could merge in a common value at high
energies)~\cite{derujula}. There was no prediction for the MM mass; a
rough estimate, obtained assuming that the classical monopole radius
is equal to the classical electron radius yields  $m_M \simeq
\frac{g^{2}m_e}{e^{2}} \simeq n \ 4700\  m_e \simeq n \ 2.4\
GeV/c^{2}$.  From 1931 searches for \textit{``classical Dirac
  monopoles"} were 
 carried out at every new accelerator using  simple setups, and
 recently also parts of large collider detectors
[3-7,20],.
\par
Electric charge is naturally quantized in Grand Unified Theories (GUT) of the 
basic interactions; they imply the existence of \textit{GUT monopoles} with 
calculable properties. The MMs would appear in the Early Universe at 
the phase transition corresponding to the breaking of the GUT group
into subgroups, one of which is U(1)~\cite{thooft}. The  MM      
mass is related to the mass  of  the X, Y carriers of the
unified interaction, $ m_{M}\ge m_{X}/G$, 
where G is the dimensionless unified coupling constant at energies E 
$\simeq m_{X}$. 
If $m_{X}\simeq 10^{14}-10^{15}$ GeV and $G\simeq0.025$,
$m_{M}>10^{16}-10^{17}$ GeV.  
This is an enormous mass: MMs cannot be produced at any man--made
accelerator, existing or conceivable. They may have been produced only
in the first instants of our  Universe and may be looked for as fossil
particles in the cosmic radiation. \par
Larger MM masses are expected if gravity is brought into the unification 
picture, and in some  SuperSymmetric models \cite{gg1}.
\par
\textit{Intermediate Mass Monopoles }(IMMs) may have been produced in later
 phase transitions in the Early Universe, when a semisimple 
gauge group
yields a U(1) group~\cite{lazaride}. IMMs with m$_M$ $\sim 10^{7} \div
10^{13}$ GeV  
may be accelerated to relativistic velocities in one galactic 
magnetic field domain. Very energetic IMMs could yield the highest
energy cosmic rays~\cite{bhatta}. 
\par
The lowest mass MM is stable, since magnetic charge is 
conserved like electric charge. Thus the poles produced in 
the Early Universe should still exist as cosmic relics; their 
kinetic energy was affected by the  
Universe expansion and by travel through galactic and 
intergalactic magnetic fields. 
\par
GUT poles are best searched for 
underground in the penetrating cosmic radiation (CR). IMMs may be
searched for at high altitude laboratories. 
 \par
\emph{Nuclearites} and \emph{Q-balls} are defined and discussed in
Section 8 \cite{nucleariti,qballs}.
\par \noindent In this lecture notes the searches for MM, nuclearites
 and Q-balls, are reviewed and
discussed.  

\section{Properties of magnetic monopoles}\label{sec:prop-mm}
The main properties of MMs  are obtained from the Dirac relation. \par 
\noindent - If $n$=1 and  the basic electric charge is that of the 
electron, then  the {\it basic magnetic charge} is 
$ g_D=\hbar c/ 2e=137e/2$. The magnetic charge is larger if  $n>1$
and  if the basic electric charge is $e/3$. 

\noindent - 
In analogy with the fine structure constant, $\alpha 
=e^{2}/\hbar c\simeq 
1/137$, the {\it dimensionless magnetic coupling constant} is 
$ \alpha_g=g^{2}_{D}/ \hbar c \simeq 34.25$; since it is $>1$
perturbative calculations cannot be used. 
\par
\noindent - {\it Energy W acquired in a magnetic field }:~  
$  W=ng_{D} B\ell=n \ 20.5$ keV/G~cm.
In a coherent galactic--length   
  ($\ell\simeq 1$ kpc,  $B\simeq 3~\mu$G), the energy gained  
by a $ g=g_{D}$ MM is $ W\simeq 1.8\times10^{11}$ GeV.
Dirac poles and IMMs in the CR may be accelerated to relativistic
velocities, GUT poles may have $10^{-4}<\beta<10^{-1}$. 
\par                 
\noindent- {\it MMs may be trapped in  ferromagnetic and paramagnetic materials}. 
\par
\noindent- \emph{Electrically
charged monopoles (dyons)} may arise as quantum--mechanical 
excitations or as M--p, M-nucleus composites.\par
\noindent- \emph{The interaction of a MM magnetic charge with a nuclear magnetic 
dipole} could lead to the formation of M--nucleus bound systems.  
Such states may exist for nuclei with 
large gyromagnetic ratios.
 \par
\begin{figure}[ht]
  \begin{center}
\includegraphics[width=0.5\textwidth]{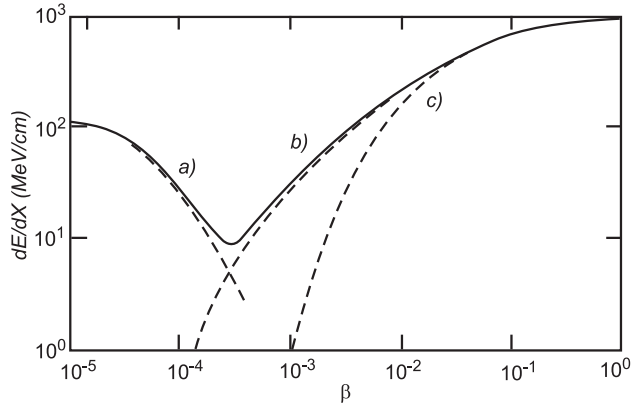}
  \end{center}
  \begin{quote}
  \caption{\small The energy losses (in MeV/cm) of $g=g_D$ MMs in
    liquid hydrogen vs ${ \beta}$: a) ionization energy loss; b) interactions
    with level crossings; c) elastic monopole--hydrogen atom scattering.} 
  \label{fig:perdita-di-energia}
  \end{quote}
\end{figure}

 \noindent- {\it Energy losses of fast poles.} 
A fast MM with magnetic charge $g_D$ and velocity $v=\beta c$ 
behaves like an electric charge 
$(ze)_{eq}=g_D\beta$, Fig.\ref{fig:perdita-di-energia}.\par
\noindent - {\it Energy losses of slow poles} ($10^{-
4}<\beta<10^{-2}$) may be due to ionization or  excitation of atoms
and molecules of the medium (``electronic'' energy loss) or to
recoiling atoms or nuclei (``atomic'' or ``nuclear'' energy
loss). Electronic energy loss predominates for $\beta>10^{-3}$. 
 \par
\noindent - {\it Energy losses at very low velocities.} 
MMs with   $v<10^{-4}c$ may lose energy in elastic collisions with atoms or 
with nuclei~\cite{derkaoui1}.\par
Fig. \ref{fig:perdita-di-energia} shows the energy losses in liquid hydrogen  
of a $g=g_D$ MM plotted vs $\beta$~\cite{gg+lp}.

\vspace{-0.7cm}
\noindent - {\it  Energy losses of MMs in celestial bodies,} for
$\beta$ $<10^{-4}$, are due to pole--atom and pole--nucleus elastic
scattering and to eddy currents. MMs may be stopped by celestial
bodies if they have:\\ 
\noindent Moon: $\beta\leq 5\times {10^{-5}}$,\quad  
Earth: $\beta \leq 10^{-4}$, \quad Sun: $\beta \leq 
10^{-3}.$\par 

\section{Monopole detectors}\label{sec:mm-det}

Monopole detectors are based on the MM properties obtained from the
Dirac relation. \par
\noindent - {\it  Superconducting induction devices are sensitive to
  MMs of any velocity} \cite{gg1}. A moving MM induces in a ring an
electromotive force and a current change ($\Delta i$). For a  coil
with N turns and inductance {\it L}: 
$ \Delta i=4\pi N ng_D/L=2\Delta i_o$,
where $\Delta i_o$ is the current change
corresponding to a change of one unit of the flux quantum of
superconductivity. This method of detection  is based on the
long--range electromagnetic interaction between the magnetic charge
and the macroscopic quantum state of a superconducting ring. \par 

\noindent - {\it Scintillation counters} 
for MMs have a threshold  
$\beta \sim 10^{-4}$, above which the light signal 
is  larger than that of a minimum ionizing particle~\cite{derkaoui1,macro1}. 

\noindent - {\it Gaseous detectors } of various types have been used. 
MACRO used a gas mixture of 73\% helium and 27\%
n--pentane~\cite{macro1}, which allows  exploitation of the
Drell~\cite{drell} and Penning effects~\cite{gg1}: a MM leaves a
helium atom in a metastable state (He*); the excited energy of the He*
is converted  into ionization of the n--pentane molecule (Penning
effect).
\par
\noindent - {\it Nuclear track detectors (NTDs).}  The formation of an etchable
 track in a NTD is related to the Restricted Energy Loss (REL), the
 fraction of the energy loss localized in a cylindrical region of few
 tens of nm diameter around the particle trajectory. It was shown that
 both the electronic and the nuclear energy losses are effective in 
 producing etchable tracks in the CR39 NTD which has a threshold at
 $z/\beta \geq 5$~\cite{cr39}; CR39 is the most sensitive NTD and it
 allows to search for MMs with $g=g_D$ for $\beta$ around $10^{-4}$
 and $>10^{-3}$, the whole  $\beta$-range of $4 \times 10^{-5}<\beta<
 1$ for MMs with $g \geq 2 g_D$~\cite{derkaoui1}. The Lexan and
 Makrofol polycarbonates are sensitive for $z/\beta \geq
 50$~\cite{barcellona}. Fig. 2
 ~~shows the calibration of CR39 and
 Makrofol NTDs with $In^{49}$ and $Pb^{82}$ relativistic lead ions and
 their fragments~\cite{CR39frag}.
 \vspace{-0.1cm}
 \begin{figure}[htb]
\hspace{-0.59cm}
   {\centering\resizebox*{17cm}{6.3cm}{\includegraphics{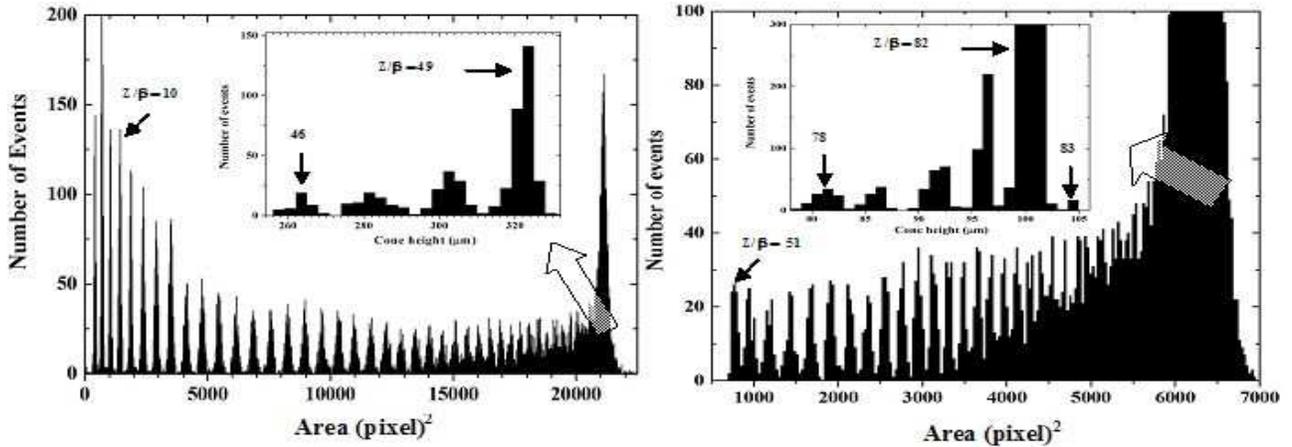}}\par}
     \label{calibr}
\vspace{-0.5cm}
   \begin{quote}
     \caption{\small Calibration of CR39 (left) and Makrofol (right)
       detectors with 158 AGeV ions ($In^{49}$ and $Pb^{82}$
       respectively) and their fragments.} 
   \end{quote}
 \end{figure}
 \vspace{-0.9cm}
 \section{Searches for ``classical Dirac monopoles''}
\begin{figure}
  {\centering\resizebox*{!}{6.5cm}{\includegraphics{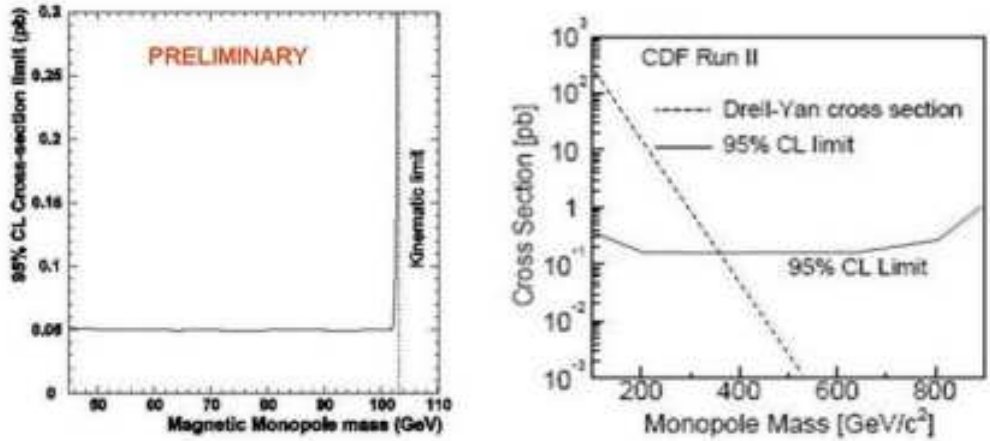}}\par}
 \vspace{-0.2cm}
  \begin{quote}
    \caption{\small \emph{Left:} 95\% CL upper limits obtained with
      the OPAL central detector at LEP2 in $e^+e^-$ collisions at
      $\sqrt s=206$ GeV. \emph{Right:} 95\% CL cross section upper
      limits vs. $m_M$ at the $p\bar p$ collider with the CDF
      experiment. If MMs are pair produced via the Drell-Yan process
      then the experiment gives the limit $m_M>360$ GeV.}\label{opalplot}
  \end{quote}
\end{figure}

\begin{figure}
 {\centering\resizebox*{!}{3.5cm}{\includegraphics{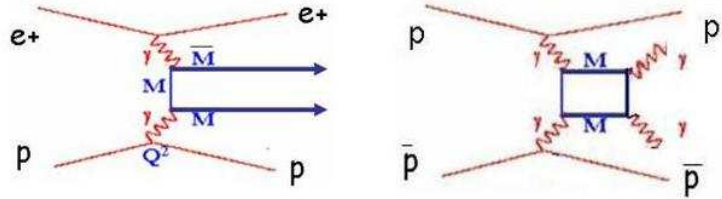}}\par}	
 \vspace{-0.2cm}
  \begin{quote}
    \caption{\small {Left:} Feymman diagram for $M \bar M$
      production in $e^{+} p$ collisions at $\sqrt s=300$ GeV,
      $e^+ p \rightarrow e^+ p M \bar M$. {Right:} indirect search 
      for monopole production in $p\bar 
    p$ collisions at Fermilab, $\bar p p \rightarrow \bar p p \gamma
    \gamma$; the two $\gamma$'s are produced via the virtual monopole
    loop \cite{abbott}
.}\label{doublegamma}
  \end{quote}
\end{figure}

 \noindent - {\it Accelerator searches.} 
 If MMs are produced at high--energy accelerators, they would 
 in general be re\-la\-ti\-vi\-stic and ionize heavily.
 Examples of \textit{direct searches} are the experiments performed 
 with scintillators or NTDs. Early experiments at the Fermilab $\overline p
 p$ collider established cross section upper limits of $\sim 2\times
 10^{-34}$~cm$^2$ for MMs with $m_M<850$ GeV~\cite{bertani}. Early
 direct searches at $e^{+}e^{-}$  colliders excluded masses up to 45 GeV
 \cite{opal}. The OPAL experiment searched for Dirac MMs in $e^+e^-$
 collisions in the $45<\sqrt s<104$ GeV range ($\sigma<5\times
 10^{-37}$~cm$^2$), Fig. \ref{opalplot} left \cite{cozzi}. The CDF
 experiment established a direct limit using some of its sub-detectors
 at the 
 $p\bar p$ Fermilab collider, see Fig. \ref {opalplot} right
 \cite{CDF}. In $e^+ p$ 
 collisions, indirect experiments placed a limit for MMs using the
 process sketched in Fig.\ref{doublegamma} left \cite{gg1}.

\par Most searches are sensitive to poles with magnetic charges $g =~n~
g_{D}/q$ with $0.5<n<5$.\par Examples of \emph{indirect searches} are those
performed at the CERN  SPS and at Fermilab: the protons interacted in
ferromagnetic or paramagnetic targets; later the targets were placed
in front of a superconducting solenoid with a field  $B>100$ kG,
large enough to extract  and accelerate the MMs, to be  detected in
scintillators and 
in  NTD sheets~\cite{gg1}. An indirect experiment performed at the
$\bar{p}p$ Tevatron collider, assumed that produced MMs could stop, be
trapped and bound in the matter surrounding a collision
region~\cite{kalbfleish}. Small Be and Al samples were passed through 
the 10 cm diameter bore of two superconducting coils, and the induced
charge  measured by SQUIDs. Limits  for $m_M<285$ GeV for
$g=g_D$ poles were published. The authors consider these experiments
as direct experiments. In our terminology they are indirect:
for their interpretations some extra hypotheses are needed, and it is
not easy to establish their validity.
\par \noindent - {\it Multi--$\gamma$ events.} 
Five peculiar photon showers found in emulsion plates exposed to
Cosmic Rays at high--altitude, are characterized by an energetic
narrow cone of tens of photons, without any incident charged
particle~\cite{multigamma}. The total energy of the photons is $\sim
10^{11}$ GeV. The small radial spread of photons  suggested a
c.m. $\gamma=(1-\beta^{2})^{-1/2}>10^3$. The energies of the photons
are too small to have $\pi^o$ decays as their source. One possible
explanation is the following, a high--energy $\gamma$--ray, with
energy  $>10^{12}$ eV, 
produced a pole--antipole pair, which suffered bremsstrahlung and
annihilation producing the final multi--$\gamma$ events.  
Searches for multi-$\gamma$ events were performed in $pp$ collisions
at the ISR at $\sqrt{s}=53$ GeV, at the $\bar{p}p$ 1.8 TeV collider
and in $e^{+}e^{-}$ collisions at LEP. The D0 experiment at FNAL
searched for 
$\gamma$ pairs with high transverse 
energies; virtual pointlike MMs may rescatter pairs of nearly real
photons into the final state via a box monopole diagram (see
Fig. \ref{doublegamma} right); a  
95\% CL limit of 870 GeV was set ~\cite{kalbfleish}.  At LEP the L3
coll. searched for $Z\rightarrow \gamma\gamma\gamma$ events; no
deviation from QED predictions was observed, setting a 95\% CL limit
of $m_M >$510 GeV~\cite{kalbfleish}. Many authors studied the effects from
virtual monopole loops~\cite{derujula,ginzburg}. Ref.~\cite{anti-d0}
criticizes the underlying theory and doubts that significant limits can
be obtained from these experiments.

In the search made by the H1 experiment at HERA, the beam pipe surrounding
the interaction region was analyzed using a SQUID magnetometer
to look for stopped monopoles; the limit is given in
the context of the so called model A based on the diagram in figure
\ref{doublegamma} left \cite{H1}.

\par \noindent - {\it Searches in bulk matter.} 
 Classical MMs could be produced  by CRs and could stop at the Earth
 surface, where they may be trapped in ferromagnetic materials. 
Bulk matter searches used hundreds of kg of material, including
 meteorites, schists, ferromanganese nodules, iron ore and others. A
 superconducting coil through which the material was passed, yielded a
 monopole/nucleon ratio in the samples $<1.2\times 10^{-29}$ at 90\% 
 CL~\cite{gg1}. \par Ruzicka and Zrelov  summarized  all searches for
 classical poles performed before 1980~\cite{ruzicka}. A more recent
 bibliography is given in Ref.~\cite{biblio}. Possible effects arising
 from low mass MMs were reported in Ref.~\cite{oscuro}.  

\section{Searches for GUT monopoles}
GUT theories of the electroweak and strong 
interations predict the existence of superheavy MMs 
produced in the Early Universe (EU) when the
GUT gauge group breaks into separate groups, one of which is 
U(1). For example one could have the following transitions:
\begin{equation}
\footnotesize
    \begin{array}{ccccc}
        {} & 10^{15}\ GeV & {} & 10^{2}\ GeV & {} \\
        SU(5) & \longrightarrow & SU(3)_{C}\times \left[
        SU(2)_{L}\times U(1)_{Y}\right] & \longrightarrow &
        SU(3)_{C}\times U(1)_{EM} \\ 
       {} & \small10^{-35}s & {} & \small10^{-9}s & {}
    \end{array}
\end{equation}
MMs would be generated as topological point defects in the GUT phase
transition $SU(5)\longrightarrow U(1)_Y$, about one pole for each
causal domain. In the standard cosmology this leads to too many poles
(\emph{the monopole  problem}). A rapid expansion of the early Universe
(\emph{inflation}) would defer the
GUT phase transition; in the simplest version of inflation the number
of generated MMs would be very small. However if there was a reheating
phase up to large enough temperatures one would have MMs produced in
high energy collisions, like 
$e^{+}e^{-}\rightarrow M\bar{M}$. 
\par The structure of a GUT MM consists in  a very small
core, an electroweak region,  a confinement region, a
fermion--antifermion condensate (which may contain 4--fermion
baryon--number--violating terms); for $r\geq$ few $fm$ a GUT pole
behaves as a point particle generating a field
$B=g/r^{2}$~\cite{picture}. 

\begin{figure}
{\centering\resizebox*{!}{6.3cm}{\includegraphics{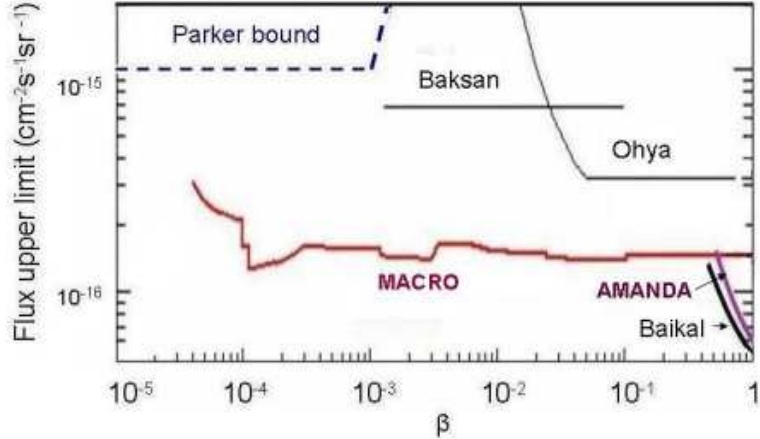}}\par}	 
\vspace{-0.5cm}
 \begin{quote}
   \caption{\small Compilation of 90\% CL direct upper
     limits vs $\beta$ for GUT  $g=g_D$ poles in the penetrating
     CR.} 
   \label{fig:global2}
 \end{quote}
\end{figure}

\par A flux of cosmic GUT MMs may reach the Earth with a velocity spectrum
in the range $4 \times 10^{-5} <\beta <0.1$, with possible peaks
corresponding to the escape velocities from the Earth, the Sun and the
Galaxy. Searches for such MMs in the  CR performed with
superconducting induction devices yielded a combined 90\%~CL limit of
$2 \times 10^{-14}~$cm$^{-2}$~s$^{-1}$~sr$^{-1}$, independent of
$\beta$~\cite{gg+lp}.
\par Direct searches were performed above ground and
underground using many types of detectors 
\cite{ruzicka, biblio, picture}.  MACRO
performed a search with liquid scintillators, limited streamer tubes and
NTDs with an acceptance of  $\sim$ 10,000 m$^2$sr for an isotropic
flux. No MM was detected. The  90\% CL flux limits, shown  in 
Fig.\ref{fig:global2} vs $\beta$  for $g=g_D$, are  at the level of
$1.4\times 10^{-16}$~cm$^{-2}$~s$^{-1}$~sr$^{-1}$ for $\beta > 4
\times 10^{-5}$~\cite{mm_macro}. The figure shows also the limits from
the Ohya~\cite{ohya}, Baksan \cite{baksan}, Baikal \cite{baikal}, and
AMANDA \cite{amanda} experiments. Adding to the MACRO limit the limit from
the SLIM experiment (described below in section \ref{IntMM}),
Fig. \ref{fig:imm1} right, over 4$\pi$, improves the MACRO limit by
about 11\%.  

\begin{figure}[htb]
 \begin{tabular}{c c}
\hspace{-0.4cm}	 
{\centering\resizebox*{!}{4.5cm}{\includegraphics{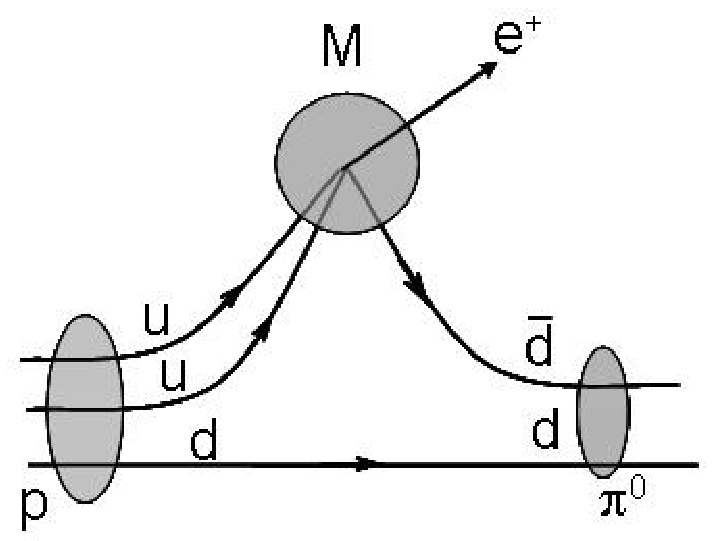}}\par}
\hspace{-2cm}	 
{\centering\resizebox*{!}{6.2cm}{\includegraphics{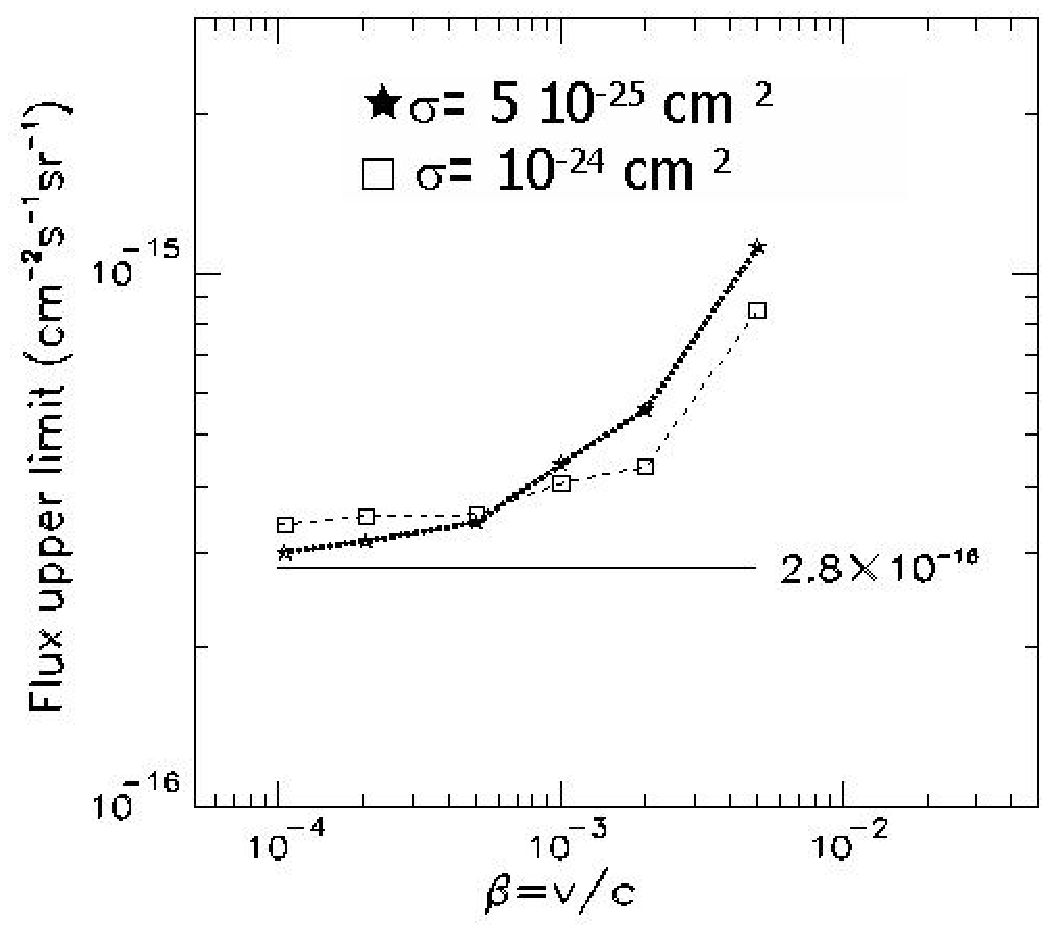}}\par}
 \end{tabular}
\vspace{-0.3cm}
 \begin{quote}
   \caption{\small \emph{Left:} Illustration of monopole catalysis of
   proton decay. \emph{Right:} MACRO flux upper limits for MM induced
   proton decay.}\label{pdecay}
 \end{quote}
\end{figure}

The interaction of the GUT monopole core with a nucleon can lead to a
reaction in which the nucleon decays (\emph{monopole catalysis of nucleon
decay}), f. e. ~~ \( M + p \rightarrow M + e^+ + \pi^0\), see
Fig. \ref{pdecay} left. The catalysis process could proceed via the
Rubakov-Callan mechanism with a $\sigma$ of the order  of the  strong
interaction cross section~\cite{rubakov}. MACRO performed a dedicated
search for nucleon decays induced by the passage of a GUT pole in the
streamer tube system. The flux limits obtained, $3-8 \times
10^{-16}$~cm$^{-2}$~s$^{-1}$~sr$^{-1}$ (Fig. \ref{pdecay} right), 
depend on the MM velocity and
on the catalysis cross section~\cite{catalisi}. Previous limits were
at levels of $10^{-15}$~cm$^{-2}$~s$^{-1}$~sr$^{-1}$~\cite{catalisi},
except the Baikal limit which is $6 \times
10^{-17}$~cm$^{-2}$~s$^{-1}$~sr$^{-1}$ for a limited $\beta$ range
around $\beta \simeq 10^{-5}$~\cite{baikal}.\par  
Indirect GUT MM searches used ancient mica samples; mica is a NTD with a
very high threshold. It is assumed that a pole passing through the Earth  
captures an Al nucleus and drags it through subterranean mica causing  
a trail of lattice defects, which survive as long as the mica is not
reheated. Only small sheets were analyzed ($13.5$ and $18$ cm$^2$),
but they should have been recording tracks for $4\div9\times 10^8$
years. The flux limits  may be at the level of $\sim ~10^{-17} ~\mbox{cm}^{-2}~
\mbox{s}^{-1} $sr$^{-1}$ for $10^{-4}<\beta<10^{-3}$~\cite{price}. 
There are several reasons why these indirect experiments might not be
sensitive.\par

\section{Cosmological and astrophysical bounds}
Rough, orders of magnitude, upper limits for a GUT monopole flux in
the CR were obtained on 
the basis of cosmological and astrophysical considerations.\par
\noindent - {\it Limit from the mass density of the universe:} 
For $m_M\sim 10^{17}$ GeV one has the limit:\par
 $F={n_Mc\over 4\pi}\beta<3\times 10^{-12}h^2_0\beta~(\mbox{cm}^{-2}\mbox{s}^{-1}
\mbox{sr}^{-1})$. It is valid for poles uniformely distributed in the
universe. If poles are clustered in galaxies the bound is
weaker~\cite{gg1}. 
\par
\noindent - {\it Limit from the galactic magnetic field (Parker bound).}
The $\sim 3\ \mu$G magnetic field in our Galaxy is probably due to the
non--uniform rotation of the Galaxy, which generates a field with a
time--scale of the order of the rotation period of the Galaxy
$(\tau\sim 10^8$ yr). An upper bound for the MM flux is obtained by
requiring that the kinetic energy gained per unit time by MMs  be less
than the magnetic energy generated by the dynamo effect:
~~ $F<10^{-15}~\mbox{cm}^{-2}~\mbox{s}^{-1}$ sr$^{-1}$~~\cite{parker}.
Taking into account the almost chaotic nature of the field, with
domains of $\ell\sim 1$ kpc, the limit becomes mass
dependent~\cite{parker}. An ``extended Parker bound'', obtained by
considering the survival of an early seed field~\cite{adams}, yields 
$ F\leq 1.2 \times
10^{-16}(m_M/10^{17}GeV)~\mbox{cm}^{-2}~\mbox{s}^{-1}~
\mbox{sr}^{-1}$. 
\par
\noindent - {\it Limit from the intergalactic (IG) magnetic field.} If 
 $B_{IG}\sim 3\times 10^{-8}~G$ ,  a more stringent  bound is obtained;
the limit is less reliable because the IG field is even less known.\par
\noindent - {\it Limits from peculiar A4 stars and from pulsars} may
be stringent, but the assumptions made are not clear (see the pulsar
PSR 1937+214)~\cite{gg1,gg+lp}. 

\section{Searches for Intermediate Mass Magnetic Monopoles (IMMs)}\label{IntMM}
IMMs may appear as topological point defects at a later time in the
Early Universe, f.e. if the GUT group yields the U(1) group of the
Standard Model in the following two steps:
\begin{equation}
\footnotesize
    \begin{array}{ccccc}
        {} & 10^{15}\ GeV  & {}& 10^{9}\ GeV & \\
        SO(10) & \longrightarrow & SU(4)\times SU(2)\times SU(2) &
        \longrightarrow & SU(3)\times SU(2)\times U(1) \\ 
        {} & \small10^{-35}s & {} & \small10^{-23}s & {}
    \end{array}
\end{equation}
\noindent This would lead to MMs with masses of $\sim 10^{10}$ GeV;
they would survive inflation, be stable, ``doubly charged'' ($g=2g_D$)
and do not catalyze nucleon decay~\cite{lazaride}. 
The structure of an IMM would be similar to that of a 
GUT MM, but the core would be larger (since R $\sim$ 1/$m_M$) 
and the outer cloud would not contain 4--fermion
baryon--number--violating terms. \par
Relativistic IMMs with masses, $10^7<m_M<10^{13}$ GeV, could be present in the
cosmic radiation, and may  be  accelerated to high $\gamma$ in one
domain of the galactic magnetic field. Thus one may look for $\beta\ge0.1$
IMMs.\par
\begin{figure}
  {\centering\resizebox*{17cm}{7cm}{\includegraphics{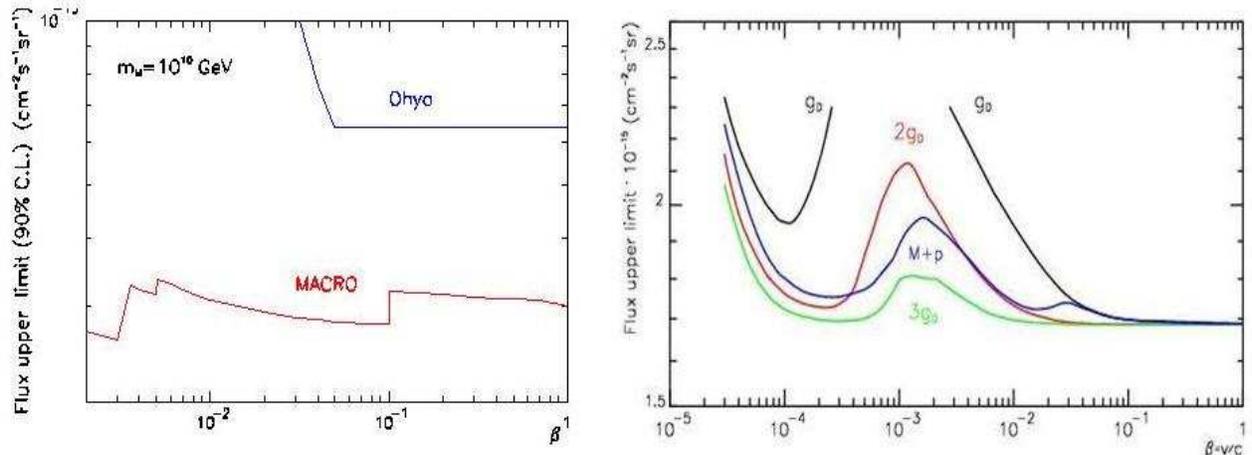}}\par}  
  \vspace{-1cm}
  \begin{quote}
    \caption{\small \emph{Left,} 90\% CL upper limits vs. $\beta$ for
      a flux of IMMs in CRs from above, with mass $m_M\sim10^{10}$
      GeV. \emph{Right:} limits for lower mass IMMs from the SLIM  
      experiment.} 
    \label{fig:imm1}
  \end{quote}
\end{figure}

  
Detectors at the Earth surface could  detect IMMs coming from above if
they have $m_M>10^5-10^6$ GeV~\cite{derkaoui1}; lower mass MMs may be
searched for with detectors located at high mountain altitudes,
in balloons and in satellites. Fig. \ref{fig:imm1} \emph{left} shows the
flux upper limits for downgoing IMMs with $m_M \sim
10^{10}GeV$ obtained by the MACRO and Ohya experiments\cite{gg+lp}.\par
\indent The SLIM experiment at the Chacaltaya High
Altitude Lab. (5260 m a.s.l.) (Bolivia) \cite{balestra} is based on
440 $m^2$ of 
CR39 and Makrofol detectors exposed for 4 years to the CR. The
detector is organized in modules of 24$\times$24 $cm^2$ each
consisting of 3 layers of CR39 (called L1, L3, and L6, respectively)
interleaved with 3 layers of Makrofol (called L2, L4 and L5,
respectively) and 1 mm Al absorber. Each module is tightly packed in an
aluminized polyethylene envelope at 1 atm of dry air to prevent the
CR39 loss in sensitivity at a reduced oxygen content in the air at the
Chacaltaya site (0.5 atmospheric pressure). SLIM is sensitive to
$g=2g_D$ IMMs in the whole range 
$4\times10^{-5}<\beta<1$ \cite{balestra}.
All CR39 was produced by the Intercast Co, Italy. An area of 351
$m^2$ of SLIM CR39 sheets 
exposed for 4 y has been etched and analysed. No candidate was
observed; the 90\% CL upper flux limits for downgoing IMMs with
$g=g_D,~2g_D,~3g_D$ and M+p are plotted in Figure \ref{fig:imm1} right,
versus $\beta$. 

\vspace{0.5cm}

\section{Nuclearites and Q-balls}
\begin{figure}
 \hspace{2cm}
  \centerline{\epsfxsize=4.2in\epsfbox{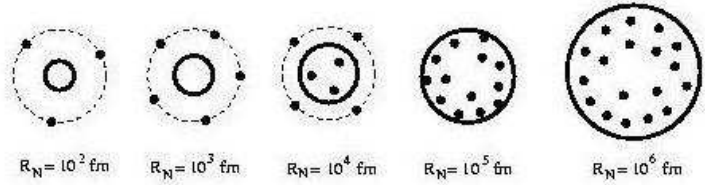}}
  \begin{quote}
  \caption{\small Sketch of nuclearite structure: the quark bag
    (radius $R_N$) and of the core+electron system; the black
    points are electrons (the border of the core+electron
    cloud for small nuclearite masses is indicated by dashed lines). For
    nuclearite masses $<~10^{9}$ GeV, the  electrons are outside and the
    core+electron system has size of $\sim 10^5$ fm; for $
    10^9<M_N<10^{15}$ GeV the $e^{-}$ are 
    partially inside the core; for  $M_N>10^{15}$ GeV all
    electrons are inside the core. \label{fig:qpict}} 
  \end{quote}
\end{figure}

\emph{Strange Quark Matter} (SQM) should consist of aggregates of \textit{u,
  d} and \textit{s} quarks in almost equal proportions; but the number of
\textit{s} quarks should be lower than the number of \textit{u} or
\textit{d} quarks and the SQM should have a relatively small positive integer
charge. The overall neutrality of SQM is ensured by an electron cloud
which surrounds it, forming a sort of atom (see
Fig.\ref{fig:qpict}) \cite{picture}.  SQM should have a constant density
$\rho_N = M_N /V_N\simeq 3.5 \times 10^{14}$~g~cm$^{-3}$, slightly
larger than that of atomic 
nuclei, and it should be stable for all baryon numbers in the range
between ordinary heavy nuclei and neutron stars (A $\sim 10^{57}$). 
Lumps of SQM with baryon number $A<10^6-10^7$ are often called
``\emph{strangelets}''; the word ``\emph{nuclearite}'' was
introduced to indicate large lumps of SQM which could be present in the
CR~\cite{nucleariti},\cite{khlopov}. SQM could have been produced shortly after
the Big Bang and may have survived as remnants; they could also appear
in violent astrophysical processes, such as neutron star collisions.  
SQM could contribute to the cold dark matter in the Universe. 
\par The main energy loss mechanism for low velocity nuclearites is
  elastic or quasi-elastic 
collisions with the ambient atoms. The energy loss  is large;
therefore nuclearites should be easily detected in scintillators and
CR39 NTDs~\cite{macro-nucl}. Nuclearites should have typical galactic
velocities, $\beta\sim10^{-3}$, and for masses larger than 0.1 g could
traverse the Earth. 
\par Most nuclearite searches were obtained as
byproducts of CR MM searches; the flux limits  are similar to those
for MMs.   
The most relevant direct flux limits for nuclearites come from three
large area experiments: the first two use CR39 NTDs; one experiment
was performed at mountain altitude (Mt. Norikura at 2770 m
a.s.l.)~\cite{nakamura}, the 2nd at the depth of $10^4$~g~cm$^{-2}$ in
the Ohya mine~\cite{ohya}; the third experiment, MACRO, at an average
depth of 3700 hg~cm$^{-2}$, used liquid scintillators besides
NTDs~\cite{gg02}.
\par Experimental limits for heavy nuclearites are at the level of
those presented in Fig.\ref{fig:global2} for GUT MMs: $\sim
1.4\times 10^{-16}~cm^{-2}s^{-1}sr^{-1}$. For Intermediate Mass
Nuclearites the limits are at the level indicated in Fig. \
\ref{fig:imm1} left, 
that is $\sim 3\times 10^{-16}~cm^{-2}s^{-1}sr^{-1}$; for slightly
smaller masses the limits of Fig.\ref{fig:imm1} right, apply: $\sim
1.7\times 10^{-15}~cm^{-2}s^{-1}sr^{-1}$ (SLIM
experiment) \cite{balestra}. For very small nuclearites, $A<8000$ the
predicted flux in the cosmic radiation is expected to increase with
decreasing mass \cite{polacchi}.
The present status of the search for galactic nuclearites is given in
ref. \cite{balestra}; the combination of the best limits come from the
AMS, SLIM and 
MACRO experiments.
\par Indirect searches could yield 
lower limits, but they are affected by several systematic uncertainties. Some
exotic cosmic ray events were interpreted as due to incident
nuclearites, f. e. the ``Centauro'' events and the anomalous massive
particles, but the interpretation is not
unique~\cite{polacchi}. Supermassive nuclearites (M$>$1 ton) passing
through Earth could possibly induce epilinear
earthquakes~\cite{nucleariti,terremoti}.    

\begin{figure}[ht]
  \centerline{\epsfxsize=3.8in\epsfbox{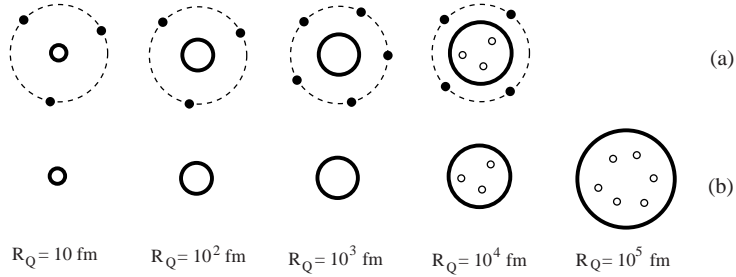}}
 \begin{quote}
   \caption{\small Sketch of the Q-ball structure: (a), SECS;
     (b), SENS. The black points represent electrons, the empty
     dots are s-electrons.} \label{qballs} 
 \end{quote} 
\end{figure}

{\it Q-balls} should be aggregates of squarks $\tilde{q}$, sleptons
$\tilde {l}$ and Higgs fields~\cite{qballs}. The scalar condensate
inside a Q-ball core has a global baryon number Q (and may be also a
lepton number). Protons, neutrons and may be electrons could be
absorbed in the condensate. 
There could exist neutral and charged Q-balls. Supersymmetric
Electrically Neutral Solitons (SENS) are generally more massive and may
catalyse proton decay. SENS may obtain a positive electric charge
when absorbing a proton in their interactions with matter yielding SECS
(Supersymmetric Electrically Charged Solitons), which have a core
electric charge, have generally lower masses and the Coulomb barrier
could prevent the capture of nuclei. SECS have only integer charges
because they are color singlets. Fig.\ref{qballs} \cite{picture} shows sketches
of SECS and SENS. A SENS which enters the Earth atmosphere could
absorb a nitrogen nucleus and would thus become a SECS with charge
$z=7$. Other nuclear absorptions may be prevented by Coulomb  
repulsion. If the Q-ball can absorb electrons at the same rate as
protons, the positive charge of the absorbed nucleus may be
neutralized by the charge of absorbed electrons. If, instead, the
absorption of electrons is slow or impossible, the Q-ball carries a
positive electric charge after the capture of the first nucleus in the
atmosphere. 
\par
\begin{figure}[h!]
  {\centering\resizebox*{!}{5cm}{\includegraphics{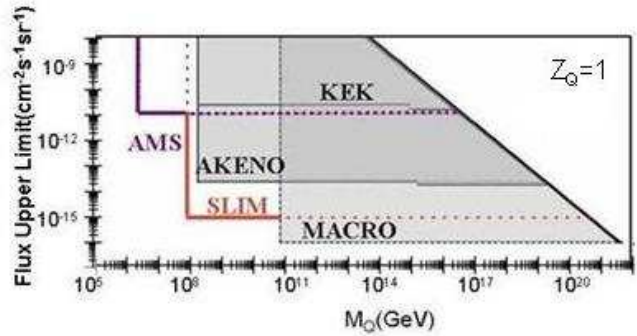}}\par}  
  \begin{quote}
    \caption{\small Flux upper limits for charged Q-balls with a net
    electric charge $Z_Q=1$.} 
    \label{upperqball}
  \end{quote}
\end{figure}

 Q-balls could be cold DM candidates. SECS with $\beta \sim 10 ^{-3}$
and  $M_Q < 10^{13}$ GeV could reach an underground detector from
above. SENS may be detected by their  continuos
emission of charged pions (estimated energy loss $\sim$100 GeV
g$^{-1}$cm$^2$); SECS may be detected by scintillators, NTDs and
ionization detectors, like those used in nuclearite and MM searches.
Fig. \ref{upperqball} shows the present status of the searches for
galactic charged Q-balls with a net charge of 1 ($Z_Q=1$) as flux
limit vs. $m_Q$. The lowest limits come from the AMS, SLIM and MACRO
experiments. 
\par We did not consider here the possibility of strongly
interacting, colored, MMs~\cite{wick}, nuclearites and Q-balls.
\vspace{-0.7cm}
\vspace{-0.6cm}

\section{Non reproducible candidates}
In the past, a number of magnetic monopole candidates and of other exotic 
events \cite{exotic} were thought to have been observed
and some results were also published in the press. But these results were not 
confirmed and most of them are now neglected.\par
\begin{figure}[h!]
 \centerline{\epsfxsize=5.3in\epsfbox{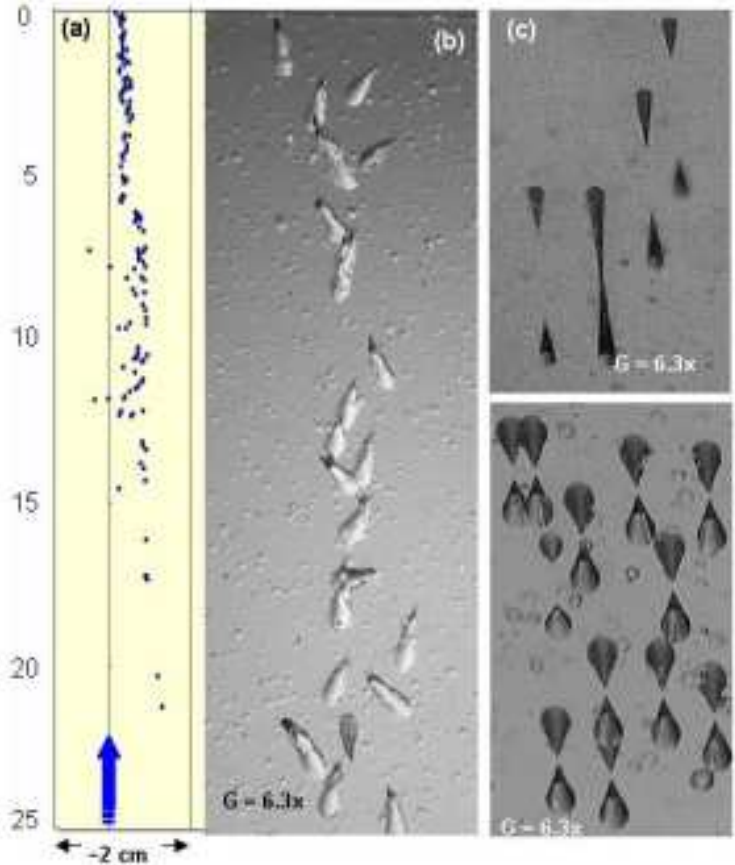}}
 \vspace{-0.5cm}
  \begin{quote}
  \caption{\small (a) Global view of the ``event'' tracks in
  the L1 layer of wagon 7408 exposed at Chacaltaya from 20-2-01 to
  28-11-05XS, (b) Microphotographs of the
  22 etched-pids at the top of Fig.11 a. (c) Normal tracks of 158 A GeV
  $Pb^{+82}$ ions and their fragments from a CERN-SPS exposure (soft
  etching), and (d) of 400 A MeV $Fe^{+26}$ ions and their fragments
  from the HIMAC accelerator, Japan (strong etching).} \label{tracce} 
\end{quote} 
\end{figure}

In 2006 the SLIM experiment faced a problem of this type when analysing the
top face of the top CR39 layer of stack 7408. We found a sequence
of many ``tracks'' (etch-pits) along a 20 cm 
line; each one of them looked complicated and very
different from usual ion tracks, see Fig. \ref{tracce} a, b. For
comparison Fig. \ref{tracce} c shows ``normal'' tracks from 158 AGeV
$Pb^{+82}$  ions and their
fragments from a CERN-SPS exposure and Fig. \ref{tracce} d shows tracks
from a 400 A MeV $Fe^{+26}$ exposure at
the HIMAC accelerator in Japan. \par
\begin{figure}[htb]
{\centering\resizebox*{!}{8.8cm}{\includegraphics{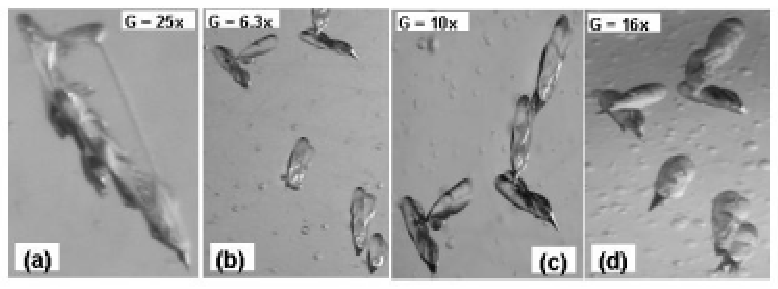}}\par} 
 \vspace{-2.5cm}
  \begin{quote}
  \caption{\small Example of ``tracks'' in the L6 layer of wagon 7410:
 (a) after 30 h of soft etching, observed
  magnification of 25x, (b) after 5 more hours of strong etching, (c)
  after 4h of more strong etching and (d) after 10h of more stong etching.}
  \label{newevent} 
\end{quote} 
\end{figure}
Since the ``event'' in the L1 sheet of module 7408 is rather
peculiar, we decided to make a thorough study of all the sheets of
module 7408, and a thorough search for similar events and
in general for background tracks in all NTD sheets in the wagons
around module 7408 (within a $\sim$ 1m distance from module 7408). We etched
``softly'' all the sheets so as to be able to follow the evolution of
the etch-pits. A second event was found in the CR39 bottom layer (top
face) of module 7410, see Fig.\ref{newevent}.  
Some background tracks in other modules were found after 30 h of
\emph{soft etching} (6N NaOH $70^{\circ}$C). We decided to further
etch ``strongly'' the 7410-L6 layer 
in short time steps (5 hr.) and to follow the evolution of the
``tracks'' by systematically making photographs
at each etching step. After additional strong etching
the ``tracks'' began looking more and more like those in the 7408-L1
layer, see Fig.\ref{newevent}b, c, d. The presence of this second
event/background and its evolution with 
increasing etching casts stronger doubts on the
event interpretation and supports a ``background'' interpretation also
of the ``tracks'' in module 7408. The background may have originated
in the fabrication of the CR39: we made different hypotheses and we
checked them with the Intercast Co. Since 1980's
we have analyzed more than 1000 $m^2$ of CR39 using different etching
conditions and we have not seen before any of the above mentioned
cases. It appears that we may have been hit by an extremely rare
manufacturing defect involving 1 $m^2$ of CR39. 

\section{Conclusions. Outlook}
\indent Direct and indirect accelerator searches for classical Dirac MMs
placed  limits for $m_M \leq 800$ GeV with specific cross section upper
values. Future improvements may come from experiments at the
LHC~\cite{moedal}.\par
\indent Many searches were performed for GUT poles in the penetrating
cosmic radiation. The 90\% CL flux limits from the MACRO experiment
are at the level of $\sim  
1.4 \times 10^{-16} $~cm$^{-2}$~s$^{-1}$~sr$^{-1}$ for $\beta \ge 
4 \times 10^{-5}$.
It may be difficult to do much better since one would require refined
detectors of considerably larger areas.
\par
Present limits on Intermediate Mass Monopoles with high $\beta$, in
the downgoing cosmic radiation are at the level of
$1.7\times 10^{-15}~cm^{-2}s^{-1}m^{-1}$. Experiments at high altitudes
and may be with neutrino telescopes should improve the situation.
\par As a byproduct of GUT MM searches some experiments obtained stringent
limits on nuclearites and on Q-balls. Future experiments at neutrino
telescopes and at high altitudes should perform searches for
nuclearites and Q-balls of smaller masses. 
\section*{Acknowledgements}
We thank the organizers of the Summer School in Puebla, Mexico, and the
discusions with several participants, in particular A. De Rujula and
O. Saavedra. The cooperation of many colleagues, in particular
S. Cecchini, M. Cozzi, M. Giorgini, G. Mandrioli, A. Margiotta, G. Sirri, V. Popa,
M. Spurio, V. Togo, is gratefully acknowledged.

\end{document}